# Optical Properties of ZnO-based Quantum Structures

Takayuki MAKINO

Photodynamics Research Center, RIKEN (Institute of Physical and Chemical Research) (519-1399, Aramaki aza Aoba, Aobaku, Sendai 980-0845)[1]

**Abstract**

The optical properties of ZnO quantum wells, which have potential application of short-wavelength semiconductor laser utilizing a high-density excitonic effect, were investigated. Stimulated emission of excitons was observed at temperatures well above room temperature due to the adoption of the lattice-matched substrates. The mechanism of stimulated emission from ZnO quantum wells is discussed in this paper.

# 1 Introduction

Recently, there have been many studies on the properties of wide gap semiconductors in response to the industrial demand for short-wavelength optoelectronic devices. Akasaki and Nakamura developed a blue light-emitting diode and a laser diode in which InGaN plays an essential role [1, 2]. All commercial semiconductor lasers including these utilize a radiative recombination of electrons and holes as the mechanism of their laser action. The threshold carrier density in these lasers is one or two orders of magnitude higher than the Mott transition density. On the other hand, the threshold in an excitonic laser is expected to be two or three orders of magnitude lower than the Mott density. Because of this expectation, interest has recently been shown in zinc oxide (ZnO), emitting light in the ultraviolet regions of the electromagnetic spectrum. In addition, its excitons are stable even at room temperature (RT) and under a high-density condition, owing to the fact of large binding energy (60 meV) as compared with those of group II-selenides or group III-nitrides.

Atomic layer control technology in laser ablation for oxides has advanced remarkably since the discovery of high-temperature superconductors. We started study of ZnO epitaxial growth by using sapphire as a substrate that had large lattice-mismatch (18%) with ZnO. Our initial research results includes: (1) an observation of the excitonic laser action using grain boundaries as a cavity [3], (2) the band gap engineering based on (Zn,Mg)O and (Zn,Cd)O [4, 5], and (3) optical characterizations of ZnO-based multi-quantum wells (MQWs) [6, 7, 8].

However, these initial studies have also revealed some problems that are unavoidable as long as a lattice-mismatched sapphire substrate is used. These thin films are indeed epitaxial but stay multicrystalline in nature having incoherent grain boundaries. Although these grain boundaries seem to be useful for observing the laser action, their electrical properties are rather poor, as represented by a typical electron concentration of $n_s \sim 10^{17}$ cm$^3$ and a typical Hall mobility of $\mu \sim 10$ cm$^2$/V s at room temperature [9]. These values are clearly inferior to those of bulk single crystals ($n \sim 10^{15}$ cm$^3$ and $\mu \sim 200$ cm$^2$/V s). In addition, unfortunately neither luminescence nor stimulated emission (SE) could be observed at room temperature in these MQWs. Thus, higher-quality quantum wells (QW) must be developed for a laser diode with a lower threshold. Furthermore, one has to eliminate a problem of the doping bottleneck because ZnO has natural tendency to be grown under fairly high-residual *n*-type [10].

We have thus adopted (0001) face of $ScAlMgO_4$ (SCAM) as a alternative substrate to

[1] electronic mail: makino@sci.u-hyogo.ac.jp, Present address: Department of Material Science, University of Hyogo, Kamigoori-cho, 678-1297, Japan

optimize the qualities of ZnO film and heterostructures; its lattice constants are $a$=0.3246 nm and $c$=2.5195 nm (Ref. [11]), which is surely lattice-matched (0.09%) to ZnO. This substrate material is regarded as a natural superlattice composed of alternating stacking layers of wurtzite-type $MgAlO_x$ and rocksalt (111)-$ScO_y$ layers and hence has a cleavage habit along the (0001) plane. A high-quality single crystal is grown through Czochralski's method. The crystal structure and a possible hetero-interface with ZnO are schematically shown in Ref. [12].

Although there is now a fairly good understanding of the basic properties of ZnO epilayers [13, 14, 15], it is only recently that the basic properties of its QWs have been studied in detail [16]. Here, we describe the optical properties of ZnO/(Zn,Mg)O MQWs grown on lattice-matched $ScAlMgO_4$ (SCAM) substrates. This paper is organized as follows. The experimental procedures are briefly described in Section 2. Improvements of ZnO thin films and the quantum confinement effects of excitons in the heterostructures are described in Section 3. The summarizing remarks are given in Section 4.

# 2 Experimentals

Samples of 10-period MQWs were grown with the laser molecular-beam epitaxy (MBE) technique. The well layer is ZnO, while the barrier layer is $Zn_{1-x}Mg_xO$. The Mg concentration dependence of the band gap energy has been given elsewhere [17]. It should be noted that the in-plane lattice mismatch between ZnO and these alloys is very small. The Mg concentration was set to 0.12 or 0.27, whereas the well width ($L_w$) ranges from 0.69 to 4.65 nm. Barrier layer thickness was fixed at 5.0 nm. The samples were grown by the "combinatorial" method, the concept of which has been explained in a review article [18]. Readers should refer to related papers in which the apparatuses developed for efficient structural and optical characterizations are described [19, 20].

# 3 Results and discussion

### 3.1 Linear optical properties

This section describes improvements of the properties of ZnO epilayers achieved by using lattice-matched SCAM substrates. An observation of the surface morphology of ZnO/SCAM epilayers revealed that the steps have a flat terrace in the scale of an atomic level and have a height of 0.26 nm (corresponding to the charge neutral unit). The crystallinity of these epilayers is almost the same level as that of the ZnO bulk as judged from their x-ray diffraction (XRD) data [12]. These films also exhibited high carrier mobility (100 $cm^2$/Vs) and low residual carrier concentration ($10^{15}$ $cm^3$). $N$-type samples were prepared through aluminum doping to examine activation efficiency of the donors, which is, as a result, about 10 times greater than that of a thin film on sapphire. The optical properties of $n$-type ZnO:Al or ZnO:Ga have been reported elsewhere [21, 22].

Another concern is the improved optical properties of our undoped films. Figure 1 shows their photoluminescence (PL) and absorption spectra measured at 4.2 K. Though the $A$-$B$ exciton splitting in ZnO is only 7 or 8 meV, these two excitonic peaks were clearly resolved for the epilayer grown on a lattice-matched substrate. On the other hand, this was not the case for the epilayer on sapphire [14] [cf. Fig. 1(a)]. The two excitonic peaks became too broad to be resolved because the damping constant of the excitons became larger due to the inferior crystallinity. The widths of the exciton absorption bands of our epilayer are similar to those of bulk crystals, as in Fig. 1(b) [23, 14]. These improvements in electrical, structural and optical properties might not be possible as long as a sapphire substrate is used. As for ZnO/buffer, the PL spectrum shows free-exciton emission ($FE_A$), $n$=2 exciton peaks and a number of bound exciton complexes. The structure of $n = 2$ excitons is also visible in corresponding absorption spectrum. The detailed assignment of the spectra is reported in the previous paper [24, 25]. It is worth mentioning that these optical properties for the ZnO/buffer are nearly identical to those of Eagle-Pitcher's single crystal. The quality of these single-crystalline ZnO epilayers satisfies the stringent requirements for being regarded as compound semiconductors.

As mentioned earlier, the MQWs on sapphire showed the following drawbacks due to the formation of rough interfaces caused by the use of lattice-mismatched substrates: (1) controllability of layer thickness is not sufficient for quantum confinement effect to be elicited in the case of $L_w \leq 1.5$ nm, and (2) PL efficiency is not sufficiently high enough to enable the observation of the RT exciton emission [6, 26]. These must be overcome for an optoelectrical device to be operable at RT [27].

The XRD patterns of the MQWs on SCAM showed Bragg diffraction peaks and clear intensity oscillations due to Laue patterns corresponding to the layer thickness [28]. This indicates a high crystallinity and a high degree of thickness homogeneity. Furthermore, an observation of the atomic force microscopy (AFM) images revealed that the surface of an MQW is composed of well-defined atomically flat terraces and steps corresponding to the charge neutral unit of ZnO. Therefore, the interface roughness in the heterostructure cannot be larger than 0.26 nm. We conclude that ZnO and MgZnO alloy layers grow in a two-dimensional growth mode on this substrate, resulting in the formation of a sharp hetero-interface between them.

Figure 2 shows PL and absorption spectra in ZnO/$Mg_{0.12}Zn_{0.88}O$ MQWs on SCAM substrates measured at 5 K with well widths ($L_w$) of 1.75 and 0.69 nm [27]. Also shown for comparison are PL and absorption spectra in a 50-nm-thick ZnO epilayer on a SCAM substrate. The blue shift of the peak energies was observed on going from $L_w$ = 1.75 nm to 0.69 nm, consistent with the quantum confinement effect. Indeed, these $L_w$'s are close to the exciton Bohr radius (≈1.8 nm). The absorption peaks ($n$=1) arise from the lowest excitonic states of well layers. The peak energies of PL were constantly located on the lower energy side of those of absorption peaks.

Figures 3(b)–(c) show the well width dependence of the peak energies of PL (closed circles) and absorption (open squares), respectively [27]. The lowest transition energy of excitons (open triangles) formed with confined electrons and holes was calculated by using the model of one-dimensional, finite periodic square-well potential. The exciton binding energy (EBE, 59 meV) is assumed to be independent of $L_w$ here, although, as will be shown later, EBE actually depends on $L_w$. The optical transition process on ZnO/$Mg_{0.12}Zn_{0.88}O$ MQW is shown in Fig. 3(a). This tendency of the $L_w$ dependence of the exciton transition energy was qualitatively reproduced by calculation. As reported by Coli, Senger and Bajaj [29, 30], incorporation of the effects of exciton-phonon interaction and quantum confinement in the calculation of the EBE, leads to the values of the excitonic transitions that agree better with our experimental data.

We here quantify the coupling constant between excitons and phonons in ZnO MQWs by analyzing the temperature dependence of the absorption spectra because the coupling constant has not been estimated quantitatively [31, 32]. Figure 4 shows the temperature dependence of the full width at half maximum (FWHM) of the excitonic absorption peaks for a ZnO epilayer (a) and for a typical MQW sample with a QW width of 1.75 nm (b). The solid line represents the fitted results based on the following equation. The temperature dependence of the FWHM can be approximately described by the following equation:

$$\Gamma(T) = \Gamma_0 + \gamma_{\rm ph} T + \Gamma_{\rm LO}/[\exp(\hbar\omega_{\rm LO}/k_{\rm B}T) - 1], \quad (1)$$

where $\Gamma_0$, $\hbar\omega_{LO}$ (72 meV), $\gamma_{ph}$, $\Gamma_{LO}$ and $k_B$ are the inhomogeneous line width at temperature ($T$) of 0 K, longitudinal optical (LO)-phonon energy, strengths of the exciton-acoustic-phonon and the exciton-LO-phonon couplings and the Bolzmann constant, respectively. Our experiments found that $\hbar\omega_{LO}$ of the MQWs is not different from the bulk value.

When we evaluate $\Gamma_{LO}$ as a function of $L_w$, $\Gamma_{LO}$ of the MQWs are smaller than those for the epilayers and monotonically decrease with decrease in $L_w$, which could be explained in terms of the quantum-confinement-induced enhancement in the EBE. Figure 5(a) shows $L_w$ dependence of EBE. As is well known, the major process that contributes to broadening of the exciton line width is scattering of 1S excitons into free-electron-hole continuum or into excited excitonic states by absorbing LO phonons. If EBE exceeds $\hbar\omega_{LO}$ (72 meV), dissociation efficiency into the continuum states is greatly suppressed compared to the case of EBE smaller than $\hbar\omega_{LO}$, giving rise to a reduced $\Gamma_{LO}$. Figure 5(b) (closed circles) shows the EBE dependence of $\Gamma_{LO}$ obtained for ZnO/$Mg_{0.12}Zn_{0.88}O$ MQWs. Here, the abscissa now corresponds to calculated binding energies by Senger and Bajaj [30]. Indeed, the EBE of some MQWs exceeds $\hbar\omega_{LO}$ [31, 33]. To explain this tendency quantitatively, we calculated contributions of both the bound and scattered electron-hole states to the line width of the lowest-lying exciton using a model of Rudin *et al.* The details of the derivation can be found in Ref. [34]. Here we just give the final result, which for the discrete part of the spectrum is

$$\gamma_{1S,dis} = \hbar\omega_{LO}(R_0/|E_0|)(\epsilon_\infty^{-1} - \epsilon_0^{-1}) \sum_{n=1}^{N} \Delta_n(q_1)|\mathrm{sgn}E_0 + n^{-2}B_1/|E_0||^{-1/2}, \quad (2)$$

where $E_0=\hbar\omega_{LO}$, $B_1$ is the binding energy of the ground-state exciton state, and $R_0$ is the Rydberg energy. $\delta_n$ is composed of a number of lengthy expressions that have been given in the Appendix of Ref. [34]. The upper limit of this summation depends on whether the $\hbar\omega_{LO}$ is larger than $B_1$ or not. In the case of $\hbar\omega_{LO}>B_1$, the calculation should be summed up to $\infty$, while $N = (1 - \hbar\omega_{LO}/B_1)^{-1/2}$ for the contrary case.

Next, the continuum part of the spectrum is discussed. The transition to the electron-hole continuum or scattering states contribute to the line width if $\hbar\omega_{LO}>B_1$. This condition is satisfied in our case only for bulk. The finally obtained expression for the linewidth parameter $\gamma_{1s,cont}$ is cited from Ref. [34] as follows:

$$\gamma_{1S,cont} = 128\hbar\omega_{LO}(\epsilon_0/\epsilon_\infty - 1)(M/\mu)\int_0^{x_0} dx \quad x(1+x)\left[1 - \exp(-2\pi/x)\right]^{-1}(F_{ee} + F_{hh} - 2F_{ee}), \quad (3)$$

We did not explicitly present $F_{ee}$, $F_{hh}$, and $F_{eh}$, which have somewhat lengthy expressions. These can also be found in the Appendix of Ref. [34]. $X_0= (\mu/m_0)^{-1/2}\ 1/2\ \varepsilon_0\ (E_0/R_0)^{1/2}$ and the integration over *x* was numerically performed. Here, $m_0$ is the vacuum electron mass and $R_0$ is 13.6 eV.

The calculated results $\gamma_{1s,disc} + \gamma_{1s,cont} = \Gamma_{LO}$ are shown with solid curves in Fig. 5(b). On the QW side, the calculated $\Gamma_{LO}$ is a monotonically decreasing function of EBE ($B_1$), which is in qualitatively good agreement with the experimental results.

It was found that the band edge luminescence in the ZnO MQWs due to radiative recombination from excitons localized by the potentials formed by the fluctuations of $L_w$ and barrier height. Our spectral assignment are based on (1) the well width dependence of Stokes shift (difference between the energies of absorption and luminescence bands), (2) the temperature dependence of PL spectra, and (3) the spectral distribution (luminescence energy dependence) of decay time constants of luminescence [27, 35, 31, 36, 37, 38]. Here, the temperature dependence of the PL spectrum in a QW in the case of magnesium composition of 0.27 is described in detail.

Figure 6 shows temperature dependence of PL spectra in a ZnO (1.75 nm)/$Mg_{0.27}Zn_{0.73}O$ MQW at temperatures of 5 to 300 K. Following a temperature rise, the PL energy of this MQW exhibited low energy shifts between 5K and 50K, blue-shifts between 50 and 200 K, and again shifts to a low energy side at temperatures higher than 200 K. The spectra obtained at 95 to 200 K had two peaks, both of which originated from recombination of localized excitons [27]. Figure 7 summarizes peak energies of the PL spectra ($E_{PL}^{pk}$) (solid circles and triangles) and the excitonic absorption energy (solid squares) as functions of a temperature for ZnO (1.75 nm)/$Mg_{0.27}Zn_{0.73}O$ and ZnO (2.79 nm)/$Mg_{0.12}Zn_{0.88}O$ MQWs. The absorption peak energies are a monotonically decreasing function of temperature [14, 31], attributed to the temperature-induced shrinkage of the energy gap. Two kinds of MQWs having different barrier heights showed significantly different temperature dependences of PL spectra. The peak energy of localized exciton PL shifts upward with the elevation of temperature due to their thermal activation. The temperature dependence, shown in Fig. 6(a) is, however, different from the abovementioned typical behavior. It should be noted that the higher PL peak position does not coincide with that of absorption spectra even at temperatures near RT; probably because of the localization energy of excitons is greater than the thermal energy. The features of excitonic spontaneous emission in the well layers are sensitively affected by the dynamics of recombination of localized exciton states, which significantly vary with temperature. Readers should refer to the original articles for details [35].

### 3.2 Nonlinear Optical Properties

As shown in Fig. 8, the spontaneous PL and absorption peaks at RT find at almost the same position each other [27, 28]. The spontaneous emission spectrum was obtained under a condition of weak He-Cd laser excitation, while the stimulated emission (SE) was measured under a condition of pulsed Nd:YAG laser (355 nm) excitation. The corresponding RT PL was not observed from the MQW grown on a sapphire substrate. We performed high-power excitation experiments for the MQWs to determine the SE characteristics. Figure 8 shows the RT SE spectra of our MQW with $x$=0.12 and $L_w$=1.75 nm. Strong and sharp emission peaks were observed at 3.24 eV above a very low threshold ($I_{th}$=17 kW/cm$^2$), and their intensity rapidly grows as the excitation intensity ($I_{ex}$) is elevated. The behavior is shown in the inset.

The $L_w$ dependences of peak energies of the absorption and SE's are summarized in Fig. 9. Similar dependence is shown in Fig. 9(c), which now concerns the threshold. The SE energy is 100 meV lower than that of the absorption peak [28, 39]. We discuss the energy difference between the SE and absorption peaks.

The temperature dependence of the SE spectrum at temperatures of 5 K to RT was estimated to clarify its mechanism. Figure 10 shows such results of the SE peak energy ($x$= 0.12, $L_w$= 3.70 and 1.75 nm). Also shown for comparison in Fig. 10(c) is the plot for a ZnO/sapphire thin film. The temperature dependence of the peak energy difference in QWs shows the same behavior as that of ZnO, suggesting the same mechanism of SE is involved for these three samples. It has been already clear that the SE band in the epilayer is what is called the $P$-line, which one of the typical phenomena of high-density exciton effects. Inelastic scattering between excitons (the Auger-like process) gives rise to the appearance of this stimulated emission band. One of the two excitons participating in the collisional events is ionized, whereas the other is recombined radiatively after collision. At sufficiently low temperatures, peak energy of the relevant SE band is lower than the resonance energy of an exciton, the energy difference of which is equal to the EBE.

Therefore, the $L_w$ dependence of EBEs can be experimentally determined from analysis of the energy position of the *P*-line. As shown in Fig. 5(a), the EBE was increased up to about 100 meV and exceeded $\hbar\omega_{LO}$ (72 meV) when $L_w$ decreased. These EBEs evaluated are in a reasonably good agreement with the theoretical values [29, 33]. Moreover, the phonon scattering process and thermo-broadening effect of excitonic line width can be controlled [cf. Fig. 5(b)] if these QW structures are used, which is favorable from the viewpoint of application.

Combinatorial concept-aided techniques adopted in the growth of our samples suppressed the variations in crystal growth conditions and hence the undesired uncertainty in the deduced spectroscopic results. In a related review article, the benefit of the combinatorial technique is described in detail [18]. For example, the well width dependence of the radiative and nonradiative recombination times of localized excitons was estimated by time-resolved photoluminescence spectroscopy [40, 36]. The well width dependence of *biexciton* binding energy was also estimated [41, 42]. In addition, optical properties of (Cd,Zn)O/(Mg,Zn)O MQWs have not ever explored so far except our work [36]. These structures are advantageous from the viewpoint of almost perfect in-plane lattice matching.

# 4 Summary

If a short-wavelength semiconductor laser (in the UV region of the electromagnetic spectrum) using exciton transitions in low-dimensional crystals (e.g., QWs) of ZnO can be produced, it is expected that threshold of the laser action will become lower than that of a laser using a conventional recombination mechanism of electrons and holes. This study has shown, through experiments on optical excitation, that it is possible to produce a laser with a low threshold (i.e., small driving current density). Laser action due to the effect of a longitudinal cavity using grain boundaries is no longer observed in ZnO epitaxial layers deposited on lattice-matched substrates, because their crystallinity has been greatly improved. Moreover, the new samples no longer have an assembly of hexagonal pillars in which corresponding resonance cavity structure has disappeared.

Moreover, a *p*-*n* junction is essential for the production of a current injection laser. A ZnO crystal usually shows *n*-type conductivity. In this paper, we have not discussed this doping bottleneck problem [10].

Such high-quality ZnO QWs deposited on lattice-matched substrates are expected to have many applications for UV optoelectronics devices. New functions of this substance have been found through research on high-quality single-crystalline thin films grown by the use of the laser-assisted molecular-beam-epitaxy. New physical properties of other oxides may also be discovered in the future.

# Acknowledgements

The quantum wells used as the target of optical characterization in our study were produced by Drs. A. Ohtomo and K. Tamura using equipment originally constructed by Dr. Y. Matsumoto, Tokyo Institute of Technology, Japan. We would like to express our gratitude to the abovementioned researchers as well as Drs. Ngyuen Tien Tuan, H. D. Sun and C. H. Chia, Institute of Physical and Chemical Research, Japan, for their assistance in all aspects of our research. Thanks are also due to Professors Y. Segawa, M. Kawasaki, H. Koinuma, and S. F. Chichibu.

# Figures

Figure 1: PL and absorption spectra measured at 5 K for ZnO films (a) on (0001) sapphire, (b) on (0001) SCAM, and (c) MgZnO buffered SCAM substrates [14]. "*A*, *B*, *C*" indicate *A*-, *B*-, and *C*-exciton bands, and "$D^0X$" shows PL of a bound exciton state. "$FE_A$" denotes free exciton emission, while "*n*" stands for the principal quantum number of the exciton series.

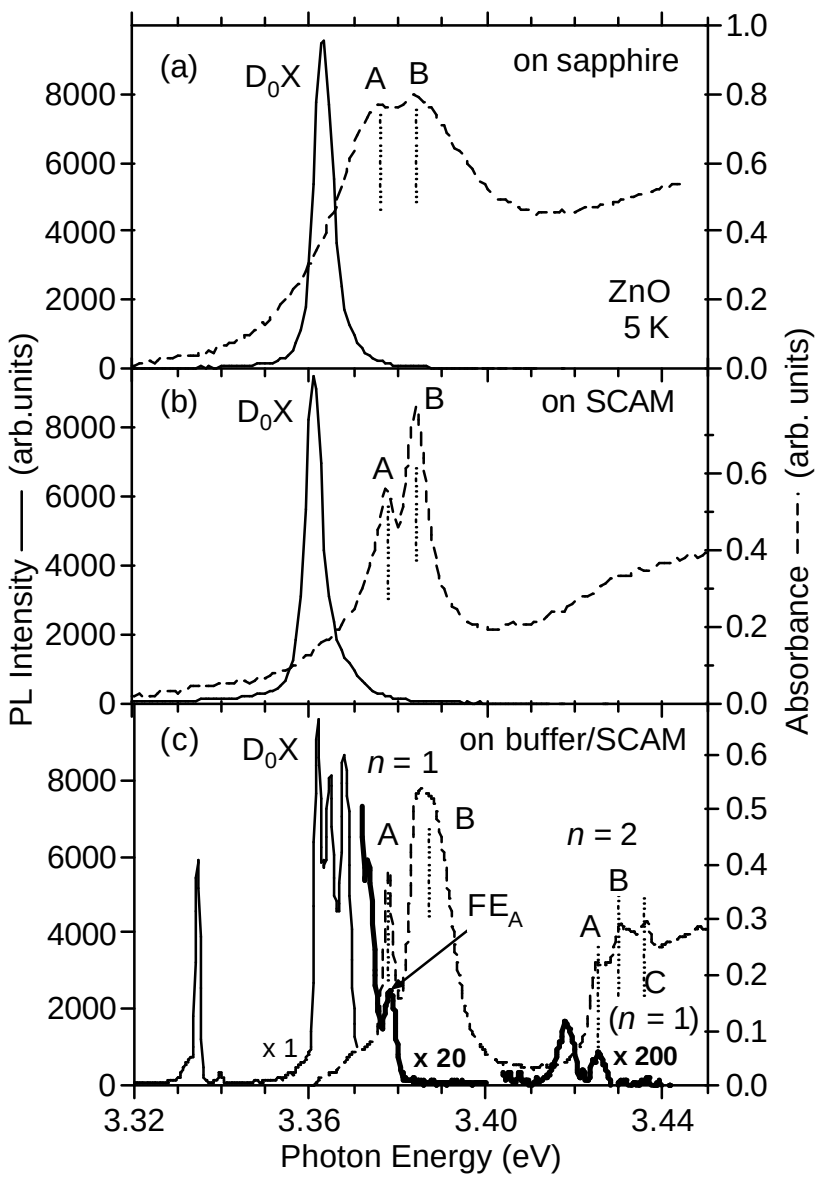

Figure 2: PL and absorption spectra obtained from ZnO/$Mg_{0.12}Zn_{0.88}O$ MQWs measured at 5 K for $L_w$ of 1.75 and 0.69 nm. A horizontal arrow shows absorption energy of barrier layers. Also shown are spectra from a ZnO film. "B+LO, B+2LO, and B+3LO" correspond to exciton-phonon complex transitions, "*n*=1" shows the lowest excitonic absorption of the well layers, and "*n*≥2" means the excited states of the exciton or higher interband transitions.

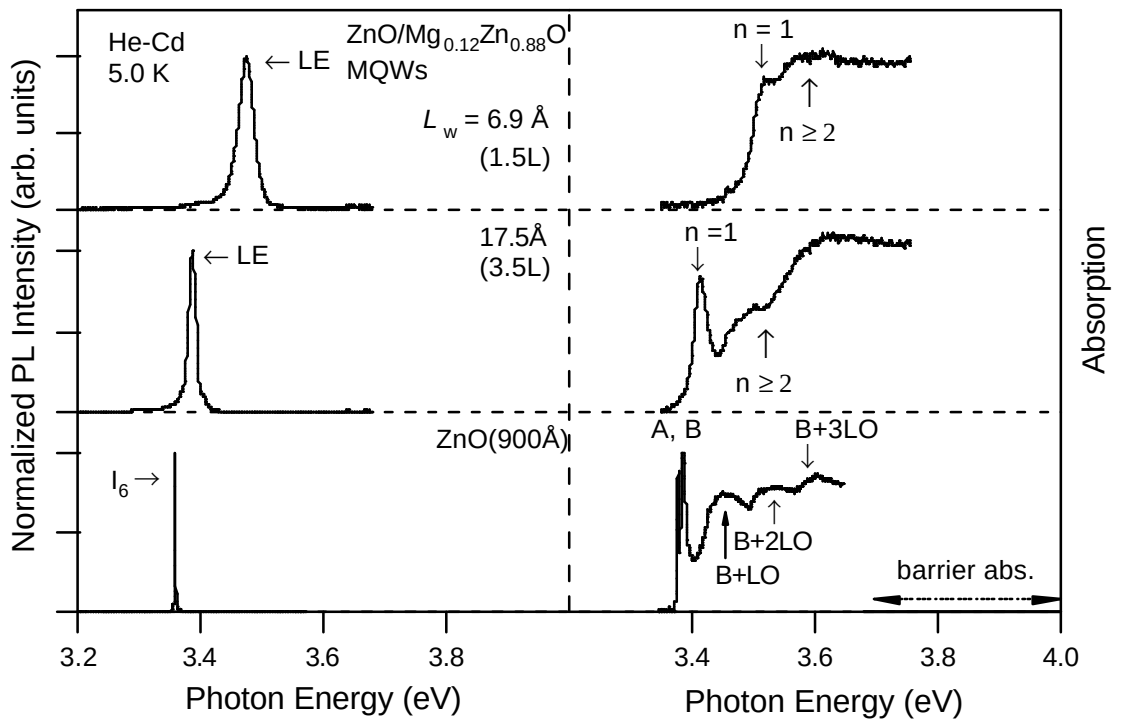

Figure 3: (a) Diagram of conduction and valence bands between barrier and well layers in a ZnO/$Mg_{0.12}Zn_{0.88}O$ MQW [6]. (b) Peak energies of PL (circles) and absorption (squares) are plotted against $L_w$ in $[ZnO/Mg_{0.27}Zn_{0.73}O]_{10}$ on SCAM substrates. Results of calculation (triangles) of the interband transition energy that include the excitonic effect are also shown. (c) Similar except that the Mg content was ≈12%.

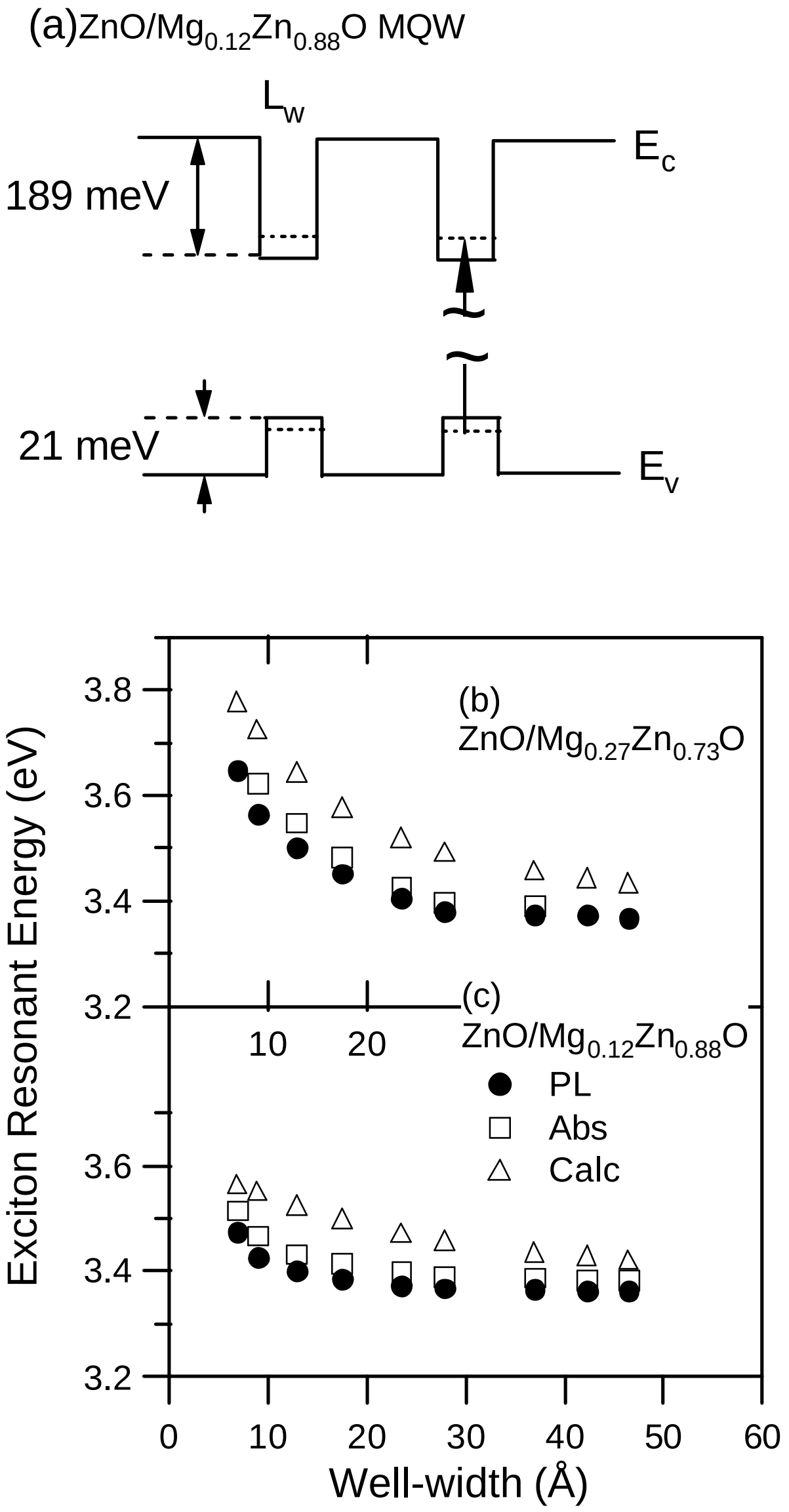

Figure 4: (a) Width (full width at half maximum, circles) of *A*- and *B*-exciton absorption bands plotted as a function of temperature. Closed circles are data of the *A*-excitons and the open circles are data of the *B*-excitons. The solid curves represent the fitting results. (b) Similar plot for the MQW with Mg concentration of 0.12 and $L_w$ of 4.65 nm [14, 31].

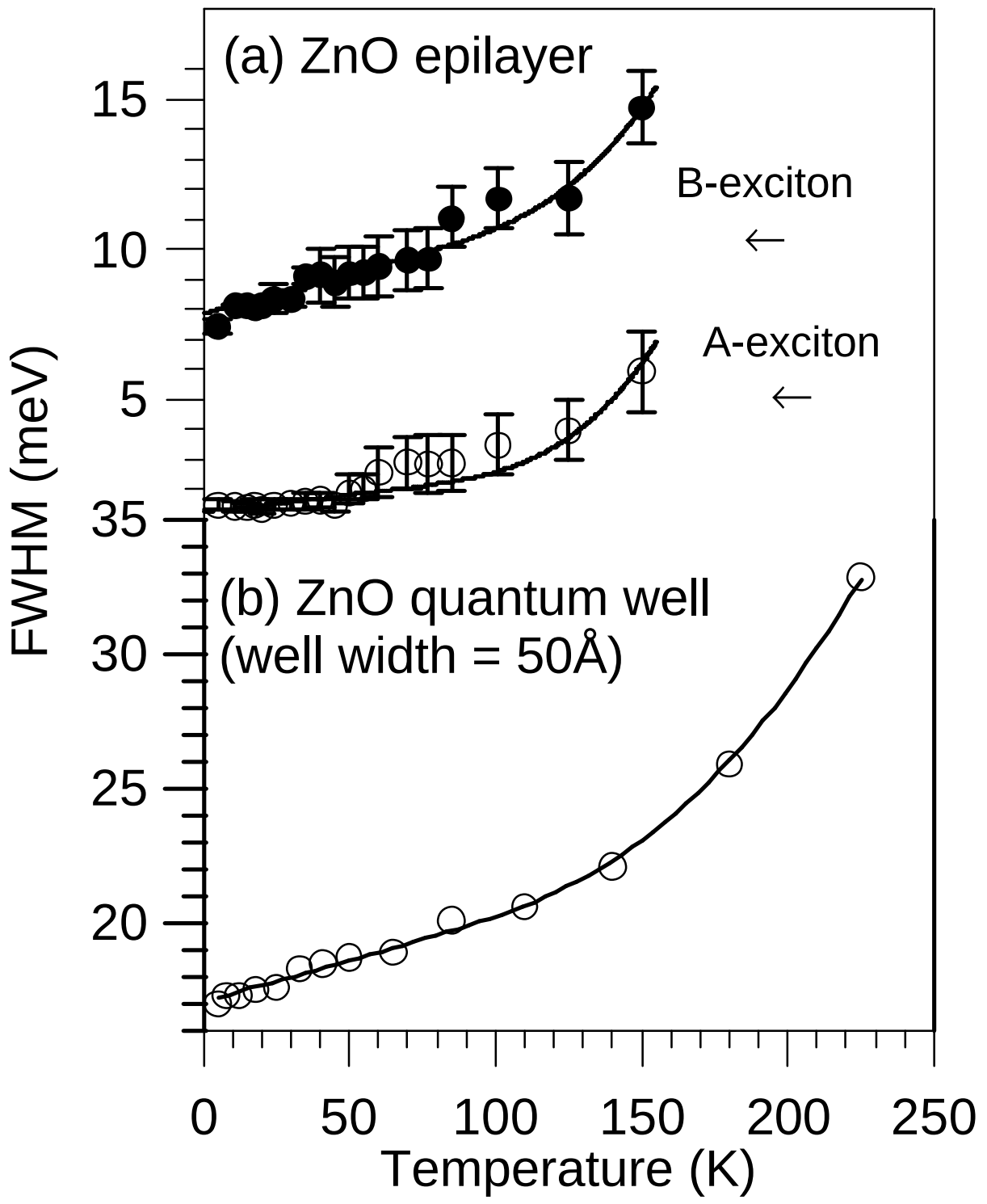

Figure 5: (a) Exciton binding energies of ZnO MQWs plotted against the $L_w$. (b) Strengths of coupling between excitons and LO phonons $\Gamma_{LO}$ (closed circles) for different EBEs [39]. The abscissa corresponds to calculated ones. Also shown by solid lines are the calculated curves for $\Gamma_{LO}$ based on Eqs. (1) and (2).

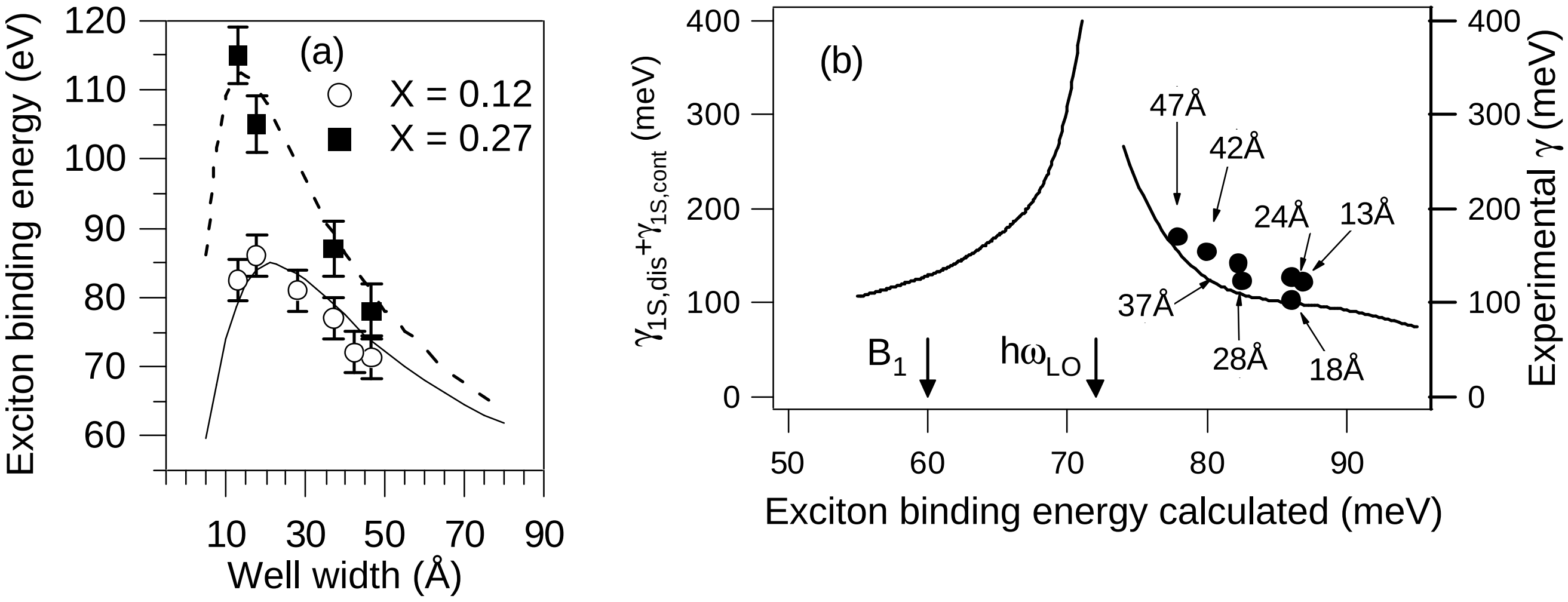

Figure 6: PL spectra in a ZnO (1.75 nm)/$Mg_{0.27}Zn_{0.73}O$ MQW at temperatures of 5 to 300 K. All of the spectra have been normalized and shifted in the vertical direction for clarity.

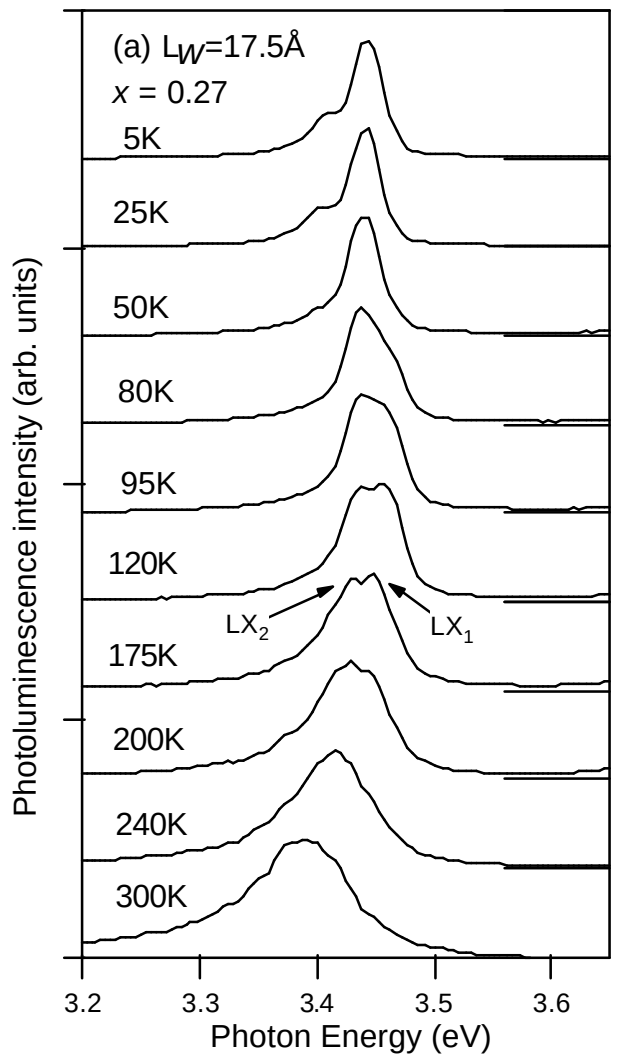

Figure 7: PL (solid circles and triangles) and absorption (solid squares) peak positions as a function of temperature in ZnO (1.75 nm)/$Mg_{0.27}Zn_{0.73}O$ (a) and ZnO (2.79 nm)/$Mg_{0.12}Zn_{0.88}O$ (b) MQWs.

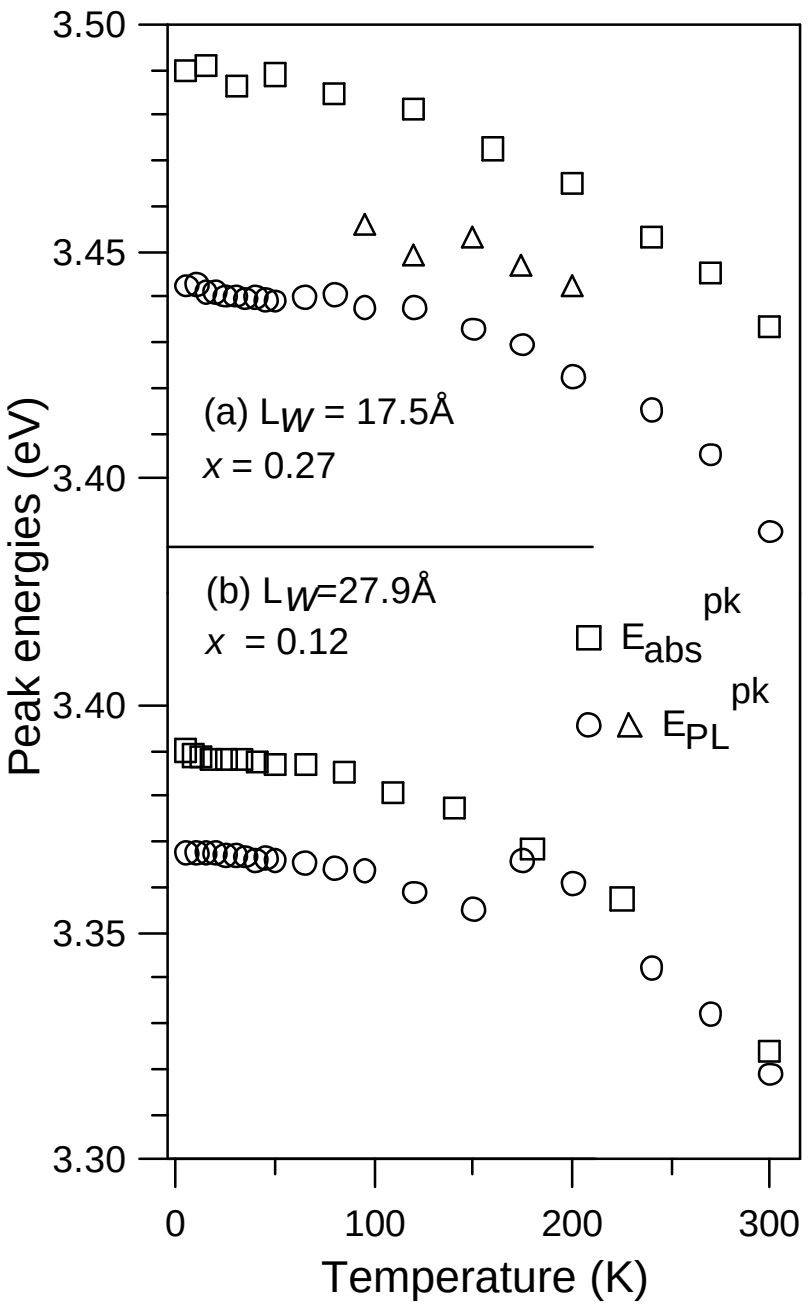

Figure 8: Excitation intensity ($I_{ex}$) dependence of the RT stimulated emission spectra obtained from a ZnO/$Mg_{0.12}Zn_{0.88}O$ MQW ($L_w$=1.75 nm) under the condition of pulsed excitation. Also shown are Spontaneous PL (dotted line) and absorption (broken line) spectra. Inset shows the integrated intensity of the stimulated emission peak as a function of $I_{ex}$. The threshold intensity ($I_{th}$) is 17 kW/cm$^2$ [28].

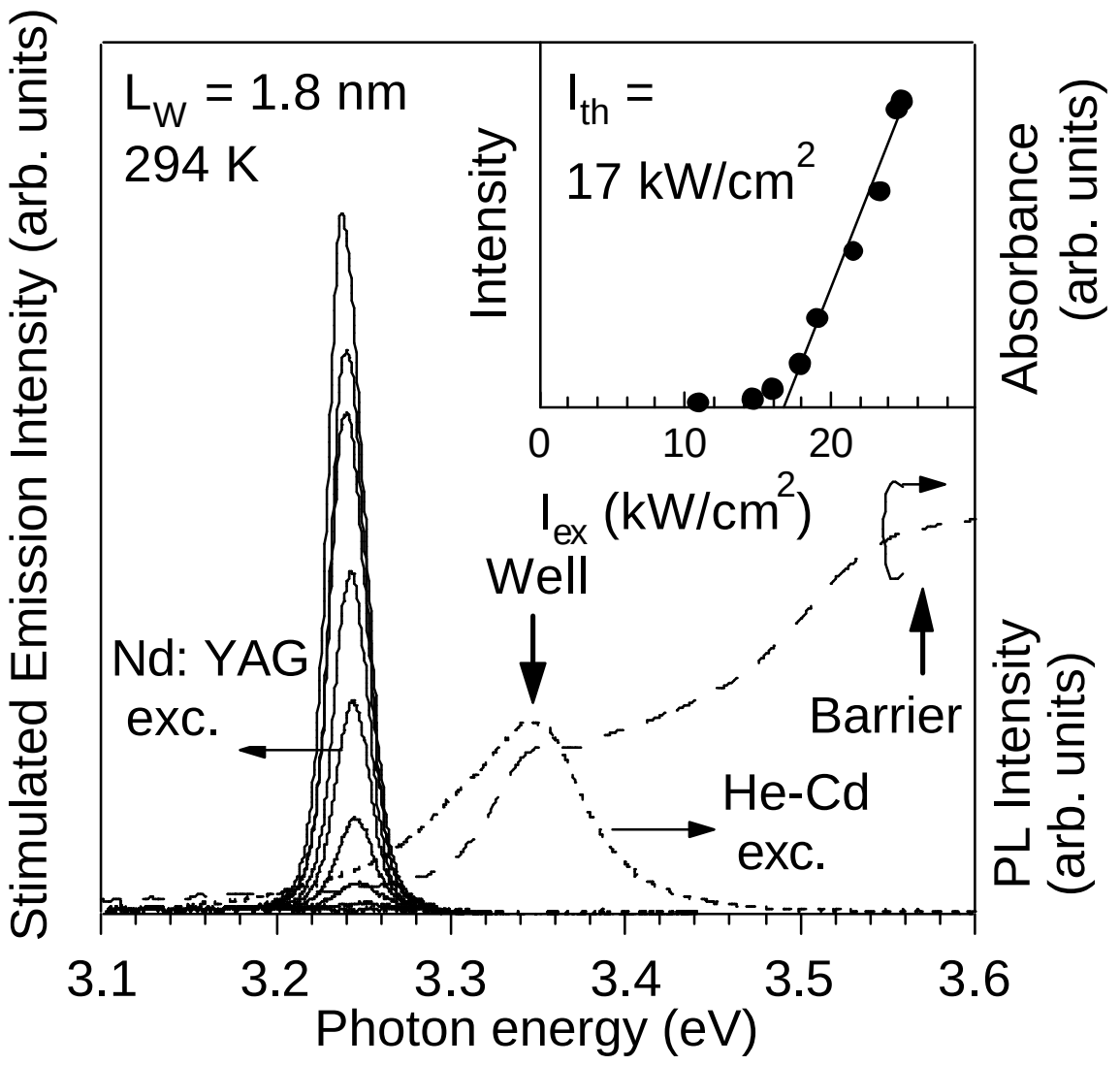

Figure 9: Optical transition energies of subband absorption (open circles) and stimulated emission (closed squares) as a function of well layer thickness ($L_w$) for ZnO/Mg$_x$Zn$_{1-x}$O MQWs with $x$=0.12 (a) and $x$=0.26 (b). Band gap energy of the barrier layers ($E_g^b$) is also shown. (c) $L_w$ dependence of the stimulated emission threshold ($I_{th}$) in MQWs with $x$=0.12 (closed circles) and $x$=0.26 (open circles). Stimulated emission did not occur for the $x$=0.26 films with $L_w$ below 1 nm since the excitation energy is lower than the absorption energy [28].

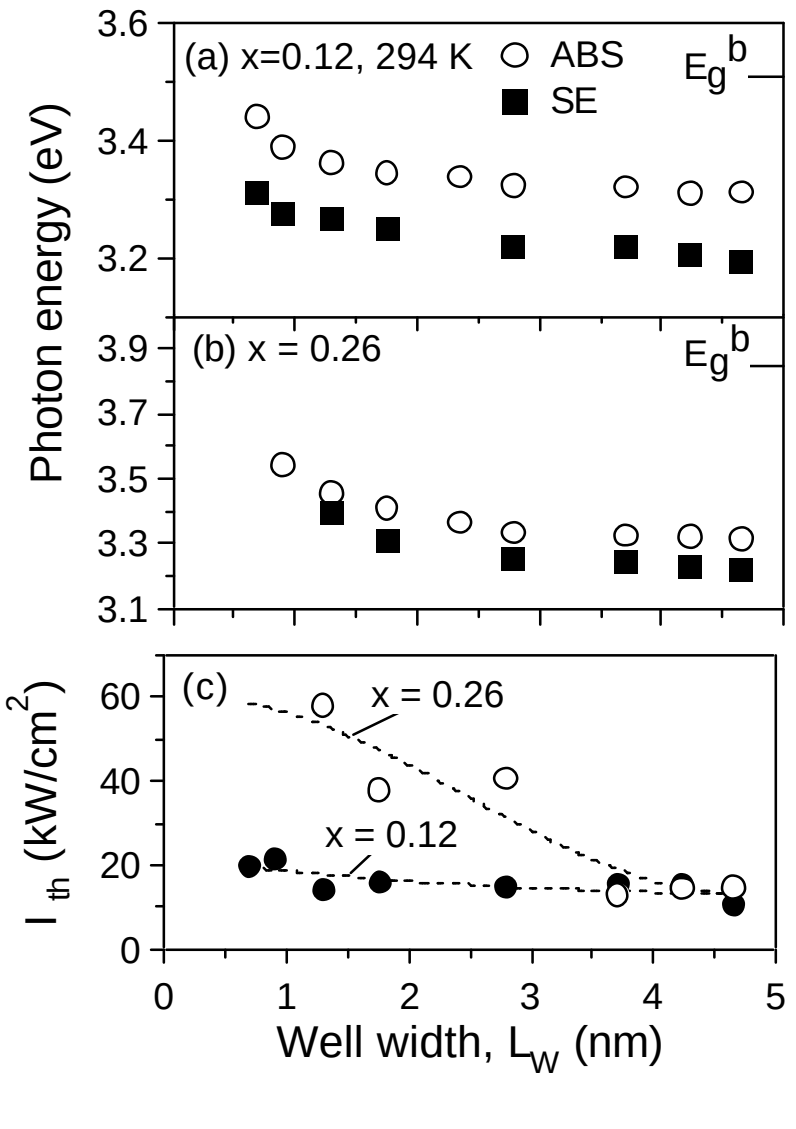

Figure 10: Temperature dependence of peak energy of the $P$ band (open circles) and free exciton energy (filled circles) in (a) a ZnO epitaxial layer and in (b, c) ZnO/Zn$_{0.88}$Mg$_{0.12}$O MQWs for two different $L_w$'s [39].

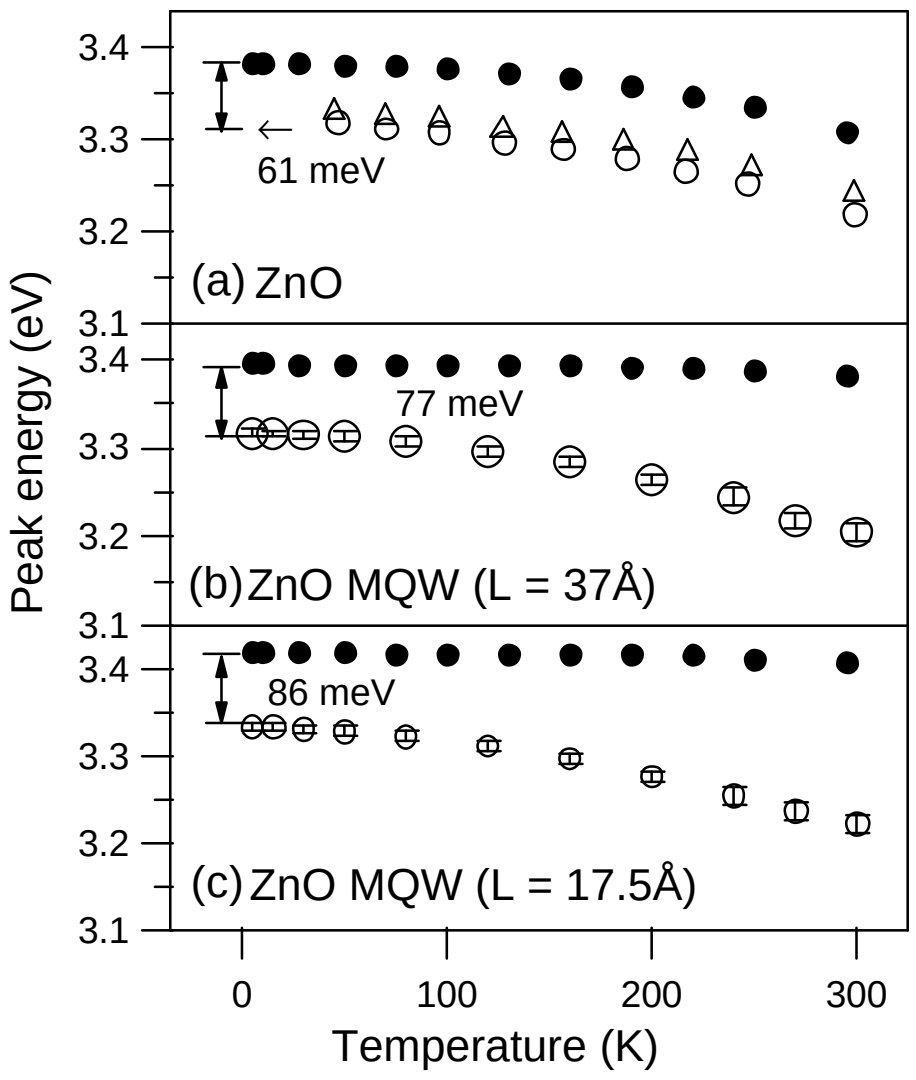